\documentstyle[12pt,eepic]{article}

\def\g{\gamma}

\def\xgo{$ x_\g^{obs}$}

\def\ETAJ{\eta^{jet}}
\def\ETJ{E_T^{jet}}

\catcode`\@=11
\long\def\@makefntext#1{
\protect\noindent \hbox to 3.2pt {\hskip-.9pt
$^{{\ninerm\@thefnmark}}$\hfil}#1\hfill}

\def\@makefnmark{\hbox to 0pt{$^{\@thefnmark}$\hss}}

\def\ps@myheadings{\let\@mkboth\@gobbletwo
\def\@oddhead{\hbox{}
\rightmark\hfil\ninerm\thepage}
\def\@oddfoot{}\def\@evenhead{\ninerm\thepage\hfil
\leftmark\hbox{}}\def\@evenfoot{}
\def\sectionmark##1{}\def\subsectionmark##1{}}

\renewcommand{\thefootnote}{\fnsymbol{footnote}}

\newcounter{sectionc}\newcounter{subsectionc}\newcounter
{subsubsectionc}
\renewcommand{\section}[1]
{\vspace*{0.6cm}\addtocounter{sectionc}{1}
\setcounter{subsectionc}{0}\setcounter{subsubsectionc}{0
}\noindent
        {\normalsize\bf\thesectionc.
#1}\par\vspace*{0.4cm}}
\renewcommand{\subsection}[1]
{\vspace*{0.6cm}\addtocounter{subsectionc}{1}
        \setcounter{subsubsectionc}{0}\noindent
        {\normalsize\it\thesectionc.\thesubsectionc.
#1}\par\vspace*{0.4cm}}
\renewcommand{\subsubsection}[1]
{\vspace*{0.6cm}\addtocounter{subsubsectionc}{1}
        \noindent
{\normalsize\rm\thesectionc.\thesubsectionc.\thesubsubse
ctionc.
        #1}\par\vspace*{0.4cm}}

\newcounter{appendixc}
\newcounter{subappendixc}[appendixc]
\newcounter{subsubappendixc}[subappendixc]

\renewcommand{\appendix}[1] {\vspace*{0.6cm}
        \refstepcounter{appendixc}
        \setcounter{figure}{0}
        \setcounter{table}{0}
        \setcounter{equation}{0}

\renewcommand{\thefigure}{\Alph{appendixc}.\arabic{figure}}

\renewcommand{\thetable}{\Alph{appendixc}.\arabic{table}}
        \renewcommand{\theappendixc}{\Alph{appendixc}}

\renewcommand{\theequation}{\Alph{appendixc}.\arabic{equation}}
        \noindent{\bf Appendix \theappendixc #1}\par\vspace*{0.4cm}}

\def\abstracts#1{{

\centering{\begin{minipage}{12.2truecm}\footnotesize\baselineskip=12pt\noindent
        \centerline{\footnotesize
ABSTRACT}\vspace*{0.3cm}
        \parindent=0pt #1
        \end{minipage}}\par}}

\renewenvironment{thebibliography}[1]
        {\begin{list}{\arabic{enumi}.}
        {\usecounter{enumi}\setlength{\parsep}{0pt}
\setlength{\leftmargin 1.25cm}{\rightmargin 0pt}
         \setlength{\itemsep}{0pt} \settowidth
        {\labelwidth}{#1.}\sloppy}}{\end{list}}

\newcounter{itemlistc}
\newcounter{romanlistc}
\newcounter{alphlistc}
\newcounter{arabiclistc}

\newcommand{\fcaption}[1]{
        \refstepcounter{figure}
        \setbox\@tempboxa = \hbox{\footnotesize
Fig.~\thefigure. #1}
        \ifdim \wd\@tempboxa > 6in
           {\begin{center}
        \parbox{6in}{\footnotesize\baselineskip=12pt
Fig.~\thefigure. #1}
            \end{center}}
        \else
             {\begin{center}
             {\footnotesize Fig.~\thefigure. #1}
              \end{center}}
        \fi}

\newcommand{\tcaption}[1]{
        \refstepcounter{table}
        \setbox\@tempboxa = \hbox{\footnotesize
Table~\thetable. #1}
        \ifdim \wd\@tempboxa > 6in
           {\begin{center}
        \parbox{6in}{\footnotesize\baselineskip=12pt
Table~\thetable. #1}
            \end{center}}
        \else
             {\begin{center}
             {\footnotesize Table~\thetable. #1}
              \end{center}}
        \fi}

\def\@citex[#1]#2{\if@filesw\immediate\write\@auxout
        {\string\citation{#2}}\fi
\def\@citea{}\@cite{\@for\@citeb:=#2\do
        {\@citea\def\@citea{,}\@ifundefined
        {b@\@citeb}{{\bf ?}\@warning
        {Citation `\@citeb' on page \thepage \space undefined}}
        {\csname b@\@citeb\endcsname}}}{#1}}

\newif\if@cghi
\def\cite{\@cghitrue\@ifnextchar [{\@tempswatrue
        \@citex}{\@tempswafalse\@citex[]}}
\def\citelow{\@cghifalse\@ifnextchar [{\@tempswatrue
        \@citex}{\@tempswafalse\@citex[]}}
\def\@cite#1#2{{$\null^{#1}$\if@tempswa\typeout
        {IJCGA warning: optional citation argument
        ignored: `#2'} \fi}}

 1
 1
 1

\font\ninerm=cmr9

\def\abbrev#1{{\sc #1}}

\def\h1{\abbrev{H1}} %
\def\cteq2l{\abbrev{CTEQ2L}} %

\def\q2{Q^2} %
\def\tq2{$\q2$} %
\def\w2{W^2} %
\def\tw2{$\w2$} %

\def\k2t{k_\perp^2}
\def\tk2t{$\k2t$}

\def\p2t{p_\perp^2}
\def\tp2t{$\p2t$}

\def\f2{F_2}
\def\tf2{$\f2$}

\def\t3{$3$}
\def\tb3{$\bar{3}$}

\def\gev2{\mbox{GeV}^2}
\def\tgev2{$\gev2$}

\def\gaeq{\,\lower3pt\hbox{$\buildrel > \over\sim$}\,}
\def\laeq{\,\lower3pt\hbox{$\buildrel < \over\sim$}\,}

\newdimen\rotdimen
\def\vspec#1{\special{ps:#1}}
\def\rotstart#1{\vspec{gsave currentpoint currentpoint translate
   #1 neg exch neg exch translate}}%
\def\rotfinish{\vspec{currentpoint grestore moveto}}%
\def\rotr#1{\rotdimen=\ht#1\advance\rotdimen by\dp#1%
   \hbox to\rotdimen{\hskip\ht#1\vbox to\wd#1{\rotstart{90 rotate}%
   \box#1\vss}\hss}\rotfinish}
\def\rotl#1{\rotdimen=\ht#1\advance\rotdimen by\dp#1%
   \hbox to\rotdimen{\vbox to\wd#1{\vskip\wd#1\rotstart{270 rotate}%
   \box#1\vss}\hss}\rotfinish}%
\def\rotu#1{\rotdimen=\ht#1\advance\rotdimen by\dp#1%
   \hbox to\wd#1{\hskip\wd#1\vbox to\rotdimen{\vskip\rotdimen
   \rotstart{-1 dup scale}\box#1\vss}\hss}\rotfinish}%
\def\rotf#1{\hbox to\wd#1{\hskip\wd#1\rotstart{-1 1 scale}%
   \box#1\hss}\rotfinish}%
\newbox\rotbox
\newbox\rottwo

\input epsf

\begin{document}

\begin{titlepage}
\begin{flushright}
CERN--TH/95--83\\
hep-ph/9506259\\
May 1995
\end{flushright}
\vspace{\fill}
\centerline{\normalsize\bf MULTIPLE HARD INTERACTIONS
IN}
\baselineskip=22pt
\centerline{\normalsize\bf\boldmath $\gamma\gamma$ AND $\gamma p$
PHYSICS
AT LEP AND HERA
\footnote{Talk given by J.M. Butterworth at Photon '95, Sheffield, U.K., April
  8--13, 1995.}}

\centerline{\footnotesize J.M. BUTTERWORTH
\footnote{butterworth@vxdesy.desy.de}}
\baselineskip=13pt
\centerline{\footnotesize\it Pennsylvania State
University, State College PA 16802. USA.}
\centerline{\footnotesize J.R. FORSHAW}
\baselineskip=13pt
\centerline{\footnotesize\it Rutherford Appleton
Laboratory, Didcot, Oxon OX11 0QX. England.}
\centerline{\footnotesize M.H. SEYMOUR}
\baselineskip=13pt
\centerline{\footnotesize\it Theory Division, CERN, CH-1211 Gen\`eve 23.
Switzerland.}
\centerline{\footnotesize J.K. STORROW AND R. WALKER}
\baselineskip=13pt
\centerline{\footnotesize\it Department of Theoretical Physics, Manchester
University, Manchester. England.}

\vspace*{0.9cm}

\abstracts{
At $e^+e^-$ and $ep$ colliders, the large
fluxes of almost on-shell photons accompanying the
lepton beams
lead to the photoproduction of jets.
As the centre-of-mass energy is increased, regions of
smaller $x$ in the
parton densities are explored and these are regions of
high parton density.
As a result, the probability for more than one hard partonic
scattering occurring in a single $\gamma \gamma$ or
$\gamma p$ collision can become significant. This effect has been
simulated using an eikonal prescription combined with the HERWIG
Monte Carlo program. The possible effects of multiple hard interactions
on event shapes and jet cross sections have been
studied in this framework at a range of energies relevant to
HERA and LEPII. The results indicate that the effects
could be significant.}

\vspace{\fill}

\noindent
CERN--TH/95--83\\
May 1995
\setcounter{footnote}{0}
\renewcommand{\thefootnote}{\alph{footnote}}
\end{titlepage}

\centerline{\normalsize\bf MULTIPLE HARD INTERACTIONS
IN}
\baselineskip=22pt
\centerline{\normalsize\bf\boldmath $\gamma\gamma$ AND $\gamma p$
PHYSICS
AT LEP AND HERA}

\centerline{\footnotesize J.M. BUTTERWORTH
\footnote{butterworth@vxdesy.desy.de}}
\baselineskip=13pt
\centerline{\footnotesize\it Pennsylvania State
University, State College PA 16802. USA.}
\centerline{\footnotesize J.R. FORSHAW}
\baselineskip=13pt
\centerline{\footnotesize\it Rutherford Appleton
Laboratory, Didcot, Oxon OX11 0QX. England.}
\centerline{\footnotesize M.H. SEYMOUR}
\baselineskip=13pt
\centerline{\footnotesize\it Theory Division, CERN, CH-1211 Gen\`eve 23.
Switzerland.}
\centerline{\footnotesize J.K. STORROW AND R. WALKER}
\baselineskip=13pt
\centerline{\footnotesize\it Department of Theoretical Physics, Manchester
University, Manchester. England.}

\vspace*{0.9cm}

\abstracts{
At $e^+e^-$ and $ep$ colliders, the large
fluxes of almost on-shell photons accompanying the
lepton beams
lead to the photoproduction of jets.
As the centre-of-mass energy is increased, regions of
smaller $x$ in the
parton densities are explored and these are regions of
high parton density.
As a result, the probability for more than one hard partonic
scattering occurring in a single $\gamma \gamma$ or
$\gamma p$ collision can become significant. This effect has been
simulated using an eikonal prescription combined with the HERWIG
Monte Carlo program. The possible effects of multiple hard interactions
on event shapes and jet cross sections have been
studied in this framework at a range of energies relevant to
HERA and LEPII. The results indicate that the effects
could be significant.}

\normalsize\baselineskip=15pt
\setcounter{footnote}{0}
\renewcommand{\thefootnote}{\alph{footnote}}

\section{Introduction}
\label{intro}
For both protons and photons,
QCD predicts a rapid increase in parton densities
at low $x$. In a naive treatment, this rise can lead to a corresponding (but
ultimately unphysical) rise with increasing energy of perturbative QCD
calculations of the jet contribution to the total cross section.
However, the large number of small $x$ partons
contributing to jet production can mean that there is a significant
probability for more than one hard scatter per $\gamma p$ or $\gamma \gamma$
interaction. The effects of multiple interactions can provide a
mechanism for taming the rise in the QCD cross section\cite{guys} in accord
with unitarity. Clearly there are implications for the hadronic final
state and in order to study these effects an eikonal model has been
implemented within the hard process generation of
HERWIG~\cite{HERWIG,Mike,jimmy}.

\begin{figure}
\begin{center}
\mbox{
\epsfxsize=2.5in
\epsfbox{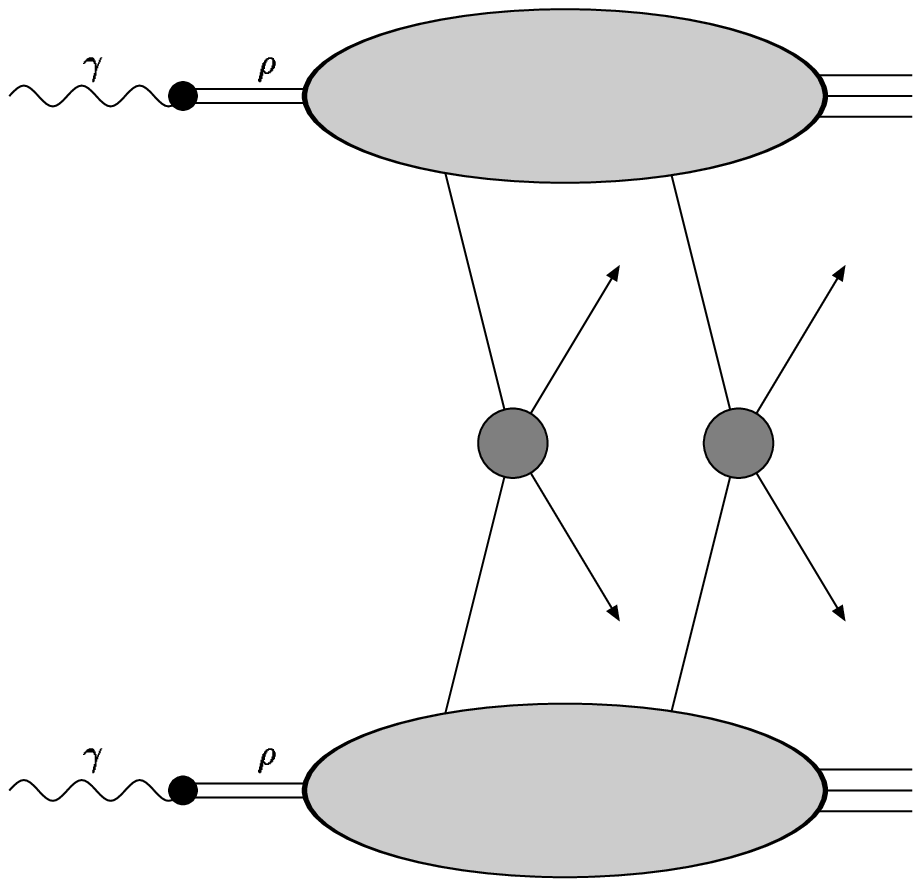}
}
\caption{\label{MI}An example of a multiple scattering in a $\gamma\gamma$
collision.}
\end{center}
\end{figure}

Multiple parton scattering, as shown in Fig.\ref{MI}, is
expected to affect jet rates.
The average number of jets per event should
be increased when partons from secondary hard scatters
are of sufficiently high
$p_T$ to give jets in their own right. In addition, lower $p_T$
secondary scatters
produce extra transverse energy in the event which
contributes to the pedestal energy underneath
other jets in the event.
Thus multiple scattering can influence jet cross
sections even when no parton from the secondary scatters is itself of a
high enough $p_T$ to produce an observable jet. By boosting the
transverse energy of jets in this way, multiple scattering
leads to an increase in jet cross sections for jets
above a certain $\ETJ$ cut,
even though the total cross section is reduced.

In the model implemented in HERWIG, the number of hard
scatters is given
by the simple probability theory formula,
\begin{center}
\#events = \#trials $\times$  event probability per
trial.
\end{center}
In the general case of a collision between particles ($A$ and $B$) at
an impact parameter $b$, the `number of trials' is the
product of the number densities of the partons in the
overlapping regions of the parent particles, i.e.
\begin{equation}
\#\hbox{trials} = \int d^2 b' \;
n_{i/A}(b',x_{A})
n_{j/B}(b'- b,x_B) = A(b) \; n_{i/A}(x_{A})
n_{j/B}(x_B),
\end{equation}
where $n_{i/A}(x_A)$ is the number density of partons of type $i$ within
particle $A$ and carrying a light cone momentum fraction, $x_A$.
We have assumed that the $x$ and $b$ dependences of the parton densities
factorize (we call $A(b)$ the area overlap function).
The `$\#$events' is the multiplicity of final state jet
pairs produced per interaction.
The `event probability per trial' is
$d\hat{\sigma}_{ij}/(P_A^{-1} P_B^{-1} \sigma_H(s))$
where $\sigma_H(s)$ is the cross section for
the production of one or more jet pairs
and $d\hat{\sigma}_{ij}$ is the
parton-parton subprocess cross section. $P_{A}$ is the
probability that
the particle $A$ actually interacts as a hadron.
For hadrons, the factor is unity, whilst for photons it is proportional to
$\alpha_{\mathrm{em}}$. In fact we take the $\rho$-dominance form
$P_{\gamma}= 4\pi\alpha_{\mathrm{em}}/f_{\rho}^2$.

The $b$ dependence of the parton density is assumed to
be given by the electromagnetic form factor of the parent particle.
By normalizing the integral of the area overlap function to
unity we can identify
\begin{equation}
A(b)d^2b \frac{\sigma^{inc}_H(s)}{P_A P_B} \equiv \chi(b,s) d^2b
\end{equation}
with the mean number of hard scatters occuring in the
element $d^2 b$
of impact parameter space (per hadronic interaction). This defines the
`eikonal' function, $\chi(b,s)$, in terms of $\sigma_H^{inc}$, the total
inclusive
cross section for the production of jets
(with $p_{T} \ge p_{Tmin}$) which is given by,
\begin{equation}
\sigma^{inc}_H(s) = \int_{p_{Tmin}^2}^{s/4}
dp_T^2 \int_{4p_T^2 /s}^{1}
 dx_A \int_{4p_T^2 x_A /s}^{1} dx_B \sum_{ij}
f_{i/A}(x_A) f_{j/B}
(x_B) \;
\frac{d\hat{\sigma}_{ij}(x_A x_B s,p_T)}{dp_T^2},
\end{equation}
where the parton distribution functions, $f_{i/A}(x_A)$, are related to the
number densities, $n_{i/A}(x_A)$, by
$f_{i/A}(x_A) = P_A n_{i/A}(x_A)$.
This inclusive cross section includes the mean
multiplicity of hard scatters
per event, i.e. $\sigma_H^{inc}(s) = \langle n_H \rangle \sigma_H(s)$.
This ensures that $\sigma_H (s)$ cannot be larger than the total cross section,
whereas, in principle, $\sigma_H^{inc}(s)$ can be.
Assuming the successive scatters to be uncorrelated, Poisson statistics then
give the probability for $m$ (and only $m$) jet
pairs to be produced:
\begin{equation}
p_m = \frac{[ \chi(b,s) ]^m}{m!} \exp[-\chi(b,s)].
\label{e:nsca}
\end{equation}

The HERWIG Monte Carlo generates the required number of
hard scatters and the associated initial and final state parton showering.
The outgoing partons and remnant jets are then fragmented to
the hadronic final state.
Two modifications are made to the simple eikonal model in the implementation.
Firstly, energy conservation is imposed, i.e.
after the backward evolution of all the hard scatters in
an event, the
energy remaining in the hadronic remnants
must be greater than
zero. Secondly, if during the backward evolution of the
first scatter,
the splitting $q\bar{q} \leftarrow \gamma$ is arrived
at before the
evolution cut off scale, the event is classified as an
`anomalous' event
and no multiple interactions are allowed (this is a conservative approach).

\section{Multiple Interactions at LEPII}

The CM energy at LEPII is sufficiently high that the
model predicts a significant proportion of multiple
scatter events. This can be seen in table \ref{nbar}, where we show the
expected mean number of hard scatters per event at a variety of colliders.
In each case, we defined a hard scatter to have $p_T \ge p_{Tmin}$
and used the MRS $D_{-}$~\cite{MRS} proton and GRV~\cite{GRV} photon
parton densities. Note that we have generated only `twice resolved'
$\gamma \gamma$ events. For these events, since the cross section falls
rapidly with increasing $p_T$, we expect that most of the multiple interactions
will have $p_T \sim p_{Tmin}$ and so their main effect will be to
increase the mean number of observed jets by boosting the underlying $E_T$ in
the event. Events which contain multiple interactions with high enough $p_T$
to be observed as jets in their own right are relatively rare, but their
observation (they appear as pairs of back-to-back jets) would provide
striking evidence in support of the existence of multiple interactions.

\begin{table}
\begin{center}
\begin{tabular}{|c|c|c|c|} \hline
 & $P_{Tmin}$ & $\sqrt{s}$ & Mean no. scatters \\
\cline{1-2} \hline
LEPII & 2.0 & 180 & 1.123 \\ \hline
LEP & 2.0 & 90 & 1.084 \\  \hline
TRISTAN & 2.0 & 60 & 1.080 \\  \hline
HERA & 3.0 & 296 & 1.04 \\  \hline
\end{tabular}
\end{center}
\caption{\label{nbar}The mean number of hard
interactions per event.}
\end{table}

\begin{figure}[ht]
\begin{center}
\mbox{
\epsfxsize=5in
\epsfbox{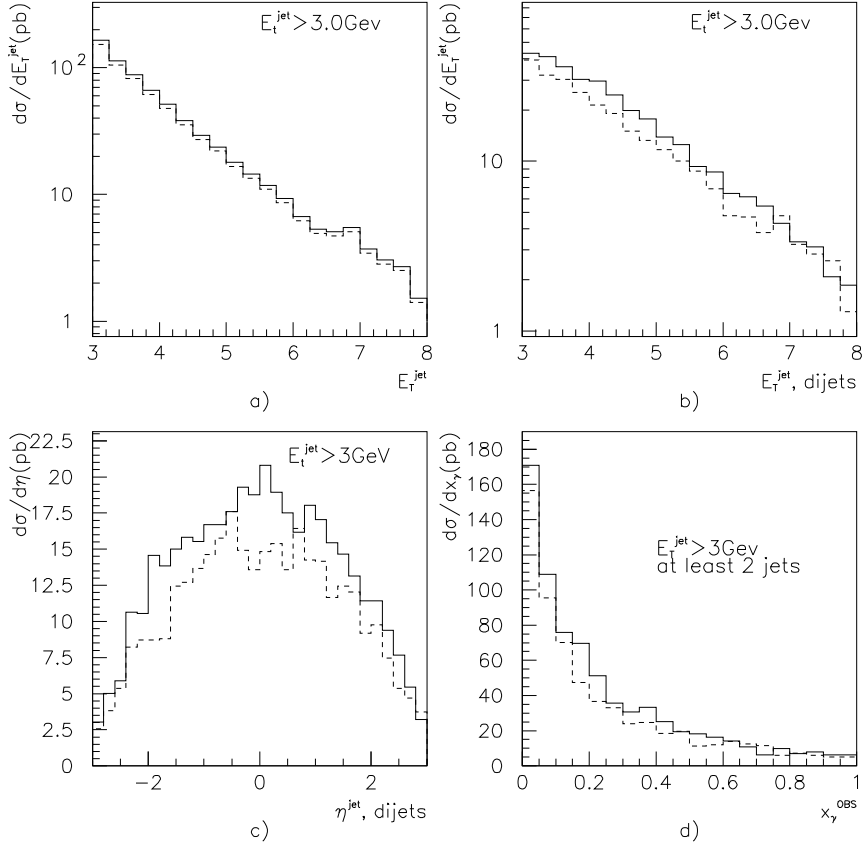}
}
\caption{\label{lepii} a) $\ETJ$ distribution for
inclusive jets b) $\ETJ$ distribution for dijets c)
$\ETAJ$ distribution for dijets d) \xgo distribution for
dijets, at LEPII energies, with multiple
scattering (solid line) and with multiple scattering
turned off (broken line).}
\vspace{-0.0in}
\end{center}
\end{figure}

Jet finding was performed on the hadronic final state
using a cone algorithm\cite{cone} with cone radius
$R=1$. Jets have $E_T \geq 3$ GeV and pseudorapidity $-3
\leq \ETAJ \leq 3$. The inclusive jet cross section, as
shown in Fig.\ref{lepii}a, is only affected by the jet
boosting, i.e. for a $p_{Tmin} = 2$ GeV, scatters from 2-3 GeV are
boosted over the $\ETJ$ cut. The effect of multiple interactions
is most significant in higher jet multiplicity rates, as can be seen in
Fig.\ref{lepii}b and c, where the dijet cross section is shown.

A variable found to be useful at HERA \cite{xgam} for distinguishing
between direct and resolved photon events is \xgo. It is
defined by
\begin{equation}
x_{\gamma}^{obs} = \frac{\sum_{jets} E_T^{jet}
e^{-\eta^{jet}}}{2E_{\gamma}}
\end{equation}
and the sum is over the two highest $E_T$ jets in the event. In leading order,
this is the fraction of the photon energy which enters the hard scatter, i.e.
direct events are peaked at $x_{\gamma}^{obs} = 1$ whilst resolved events have
$x_{\gamma}^{obs}< 1$. The same variable can be used in $\gamma \gamma$
collisions where twice resolved events populate the lower $x_{\gamma}^{obs}$
region. In Fig.\ref{lepii}d, the cross section is plotted as a
function of \xgo. It
can be seen that the effect of multiple scattering is
greater in the low \xgo region.
Multiple scatterings are more likely to occur here as it is in this region
that the higher parton densities occur; also the energy conservation
constraint is less restrictive. We have not included once resolved and
direct $\gamma \gamma$ events which will populate the high \xgo region and
should be relatively unaffected by multiple interaction.

\section{Multiple Interactions at HERA}

For HERA, the minimum transverse momentum of a hard
scatter is set to $p_{Tmin}= 3$ GeV. A cut on the $\gamma p$ CM
energy was made, i.e. 114 GeV $\leq \sqrt{s} \leq$  265 GeV, similar to those
usually made by the experiments. For these choices the mean
number of hard scatters per event was found to be 1.04.

\begin{figure}
\vspace{-1.0in}
\begin{center}
\mbox{
\epsfysize=6.5in
\epsfbox{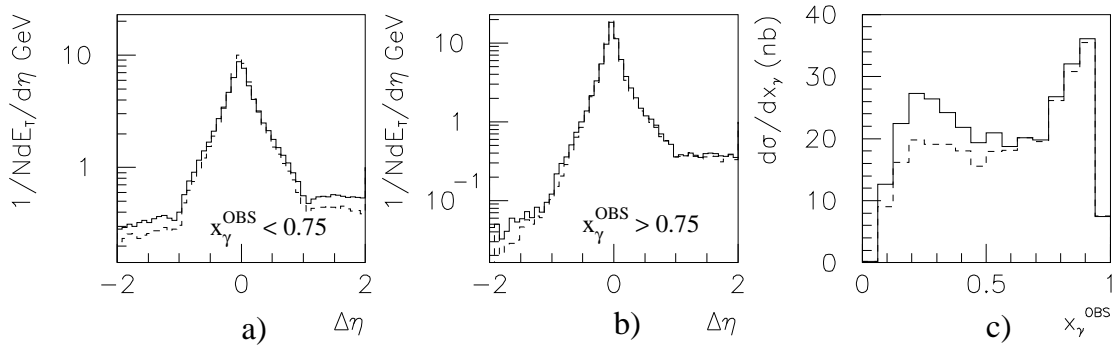}
}
\vspace{-4.0in}
\caption{\label{hera}The $E_T$ jet profile in $\eta$,
for a) \xgo$< 0.75$ and b) \xgo$\ge 0.75$. c) \xgo
distribution for dijets, at HERA energies, with direct contribution.
Including multiple scattering (solid line) and with
multiple scattering turned off (broken line).}
\end{center}
\end{figure}

Jet finding was again performed with cone radius $R=1$. Jets
have $E_T \geq 6$ GeV and pseudorapidity $-2 \leq \ETAJ
\leq 2$. The additional $E_T$ can be seen directly in
the jet profile, Fig.\ref{hera}a, where the $E_T$ in the
jets is plotted against $\eta$ relative to the jet axis.
The pedestal energy is increased with the inclusion of
multiple interactions. The \xgo cut isolates the resolved interactions, and as
expected multiple interactions have no effect in the direct case,
Fig.\ref{hera}b. The extra $E_T$ enhances the inclusive
jet rate around $\ETAJ = 1$, and an increased
sensitivity to multiple scattering can be
seen in the $\ETAJ$ and $\ETJ$ distributions in dijet
events, which are shown in ref.4. An enhancement at high $\ETAJ$ and low $\ETJ$
is
seen. The \xgo distribution, Fig.\ref{hera}c, has the
direct contribution generated using HERWIG included,
hence the rise at \xgo$\sim 1$. The inclusion of
multiple scattering has a significant effect on the
lower \xgo region, as in the $\gamma \gamma$ case.

\section{Conclusions}

The effect on the hadronic final state of multiple
parton scattering in $\gamma \gamma$ and $\gamma p$
interactions has been simulated by interfacing an
eikonal model of multiple parton interactions with HERWIG.
Multiple scattering must occur at some CM energy, in
order that unitarity is
not violated. Our model indicates that the effect of
multiple scattering is
significant at both HERA and LEPII energies. For
reasonable experimental cuts, the inclusion of multiple
scattering leads to significant changes in inclusive and
dijet cross sections which should be understood before attempting
to unfold to parton distribution functions.

This work was funded in part by the US NSF.

\end{document}